# CRITICAL EXPONENTS OF SMALL ONE-DIMENSIONAL ISING MAGNETIC

D. V. Spirin, V. N. Udodov

Within the framework of a generalized Ising model, a one-dimensional magnetic of a finite length with free ends is considered. The correlation length exponent v, dynamic critical exponent z of the magnet is calculated taking into account the next nearest neighbor interactions and the external field.

## Introduction

Within recent years, quasione-dimensional magnetics with chain structure and exhibiting one-dimensional properties at certain temperatures ($RbCoCl_3$, $CsCoBr_3$, Magnus green salt) have been intensively studied [1–13]. Of special interest are non-equilibrium processes taking place within the critical temperature interval, which are characterized y critical indices such as dynamic critical exponent $z$ [12]. Due to significant difficulties encountered in the experimental investigations (e.g., measurement of $z$ [12, 13]), a natural solution to this complex problem would be modeling of those non-eqilibrium processes.

This work addresses non-equilibrium processes in one-dimensional magnetic. Using the Monte Carlo method, an equilibrium correlation length critical exponent and the dynamic critical exponent $z$ are calculated for a finite-size magnetic.

## 1. Model of small one-dimensional Ising magnetic

A one-dimensional magnetic is a system consisting of a straight chain of $N$ atoms (sites), with every atom possessing its magnetic moment.

Let us assume that the energy of interactions between the nearest neighbors is positive $J_1 > 0$ and the initial state is antiferromagnetic (or ferrimagnetic).

With time, the system will transfer into ferromagnetic state, that is, it will undergo a non-equilibrium antiferromagnetic (ferromagnetic) – ferromagnetic phase transition. The Monte Carlo method (of numerical modeling) offers a possibility to study this transition. There are a few algorithms to implement the Monte Carlo method, one being the Metropolis algorithm [14]. A widely used model of onedomensional magnets is the Ising model [14–16], where the spins are located in a one-dimensional lattice, and the energy of this magnet is equal to

$$\frac{E}{J_1} = -\sum_{i=1}^{N-1} S_i S_{i+1} - J_2 \sum_{i=2}^{N-2} S_i S_{i+2} - H \sum_{i=1}^{N} S_1 , \qquad (1)$$

where $J_2$ is the relative energy of interactions of the next nearest neighbors, $N$ is the number of atoms (sites) in the magnetic, $i$ is the number of site, $S_i$ (acquires the values +1 or −1) is the projection of a dimensionless vector onto the axis along which the external magnetic field strength is directed, and $H = H_x$ is the dimensionless projection of the magnetic strength onto this axis.

The relaxation time for a finite one-dimensional model in a critical region is a power function of the number of sites [14]

$$\tau \propto N^z . \qquad (2)$$

It is evident from Eq. (2) that $z$ characterizes the dependence of the relaxation time on the size of a system.

Having calculated a few values of $\tau$, we can, using the method of linear extrapolations, find the values of the exponent $z$.

A similar method was used to obtain the critical exponent of the correlation length $\nu$ [15]

$$\xi \propto (T - T_c)^{-\nu}. \tag{3}$$

In what follows, we will use the reduced temperature $\theta = \dfrac{kT}{J_1}$ that will be denoted by $T$.

## 2. Results of calculations

We have calculated the relaxation times $\tau$ of the magnet under study as a function of the number of sites $N$ for different values of the magnetic field strength projection $H$ (Fig. 1) It is evident from the figure that for sufficiently large $N$ the external magnetic field does reduce the relaxation time. Note that when the sign at the external field strength is changed, there is no change in the dependence of $\tau$ on $N$.

The $z$ exponent was calculated from two close values of $N$, with $<N>$ being their simple average. The dynamic critical exponent $z$ in the zero field is decreasing with increasing $N$ (from 3.16 for $N = 3$ to 2.03 for $N = 10$) (Fig. 2, curve $1$). When the field is included into consideration, the dependence of $z$ on the average number of sites $<N>$ becomes weaker. The values of $z$ for the increasing external magnetic field strength $H$ decrease. For large $N$, the $z$ exponent is a decreasing function of the number of sites (Fig. 2).

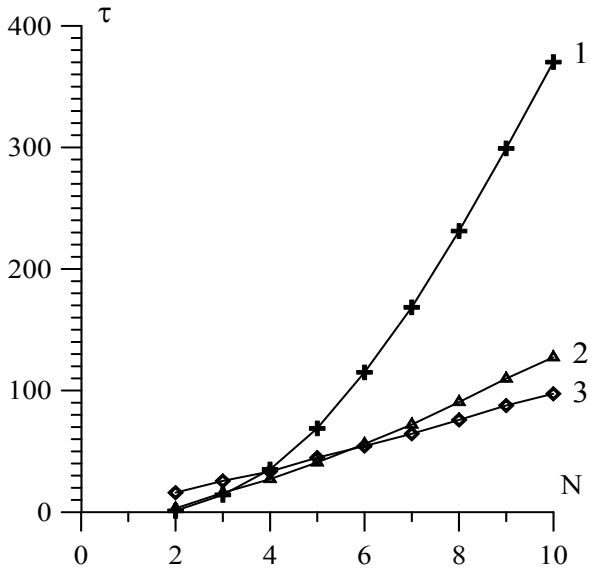

Fig. 1. Dependence of the relaxation time $\tau$ on the number of sites $N$: $T = 0.1$; $J_2 = 0$; 1) $H = 0$; 2) $H = 0.5$; 3) $H = 1$

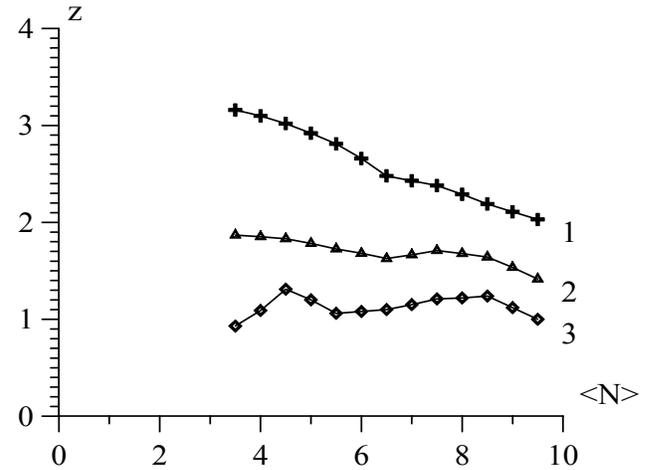

Fig. 2. Dependence of the critical dynamic exponent $z$ on the average number of sites $<N>$: $T = 0.1$; $J_2 = 0$; 1) $H = 0$; 2) $H = 0.5$; 3) $H = 1$

For positive values of the next nearest neighbor interaction, an increase in the interaction energy $J_2$ results in a decrease in the relaxation time (Fig. 3) in the same fashion as does the presence of an external magnetic field (Fig. 1). The behavior of $z$ with increasing $J_2$ is different from that for the increasing external field (Fig. 4), where curve *1* corresponds to the values of $z$ for the same parameters as those in Fig. 2. For the average number of sites smaller than six, the values of the $z$ exponent, taking into account the next nearest neighbor interaction energy, are smaller than they are for $J_2 = 0$, while at $<N>$ more than six, the situation is the opposite. It is also evident from Fig. 4 that for considerably large $N$, the dynamic critical exponent $z$ is decreasing with increasing number of sites. The energy $J_2$ used here is the ratio of the next nearest neighbor interaction energy to that of the nearest neighbor interaction; hence in Figs. 3 and 4 these energies have the same sign.

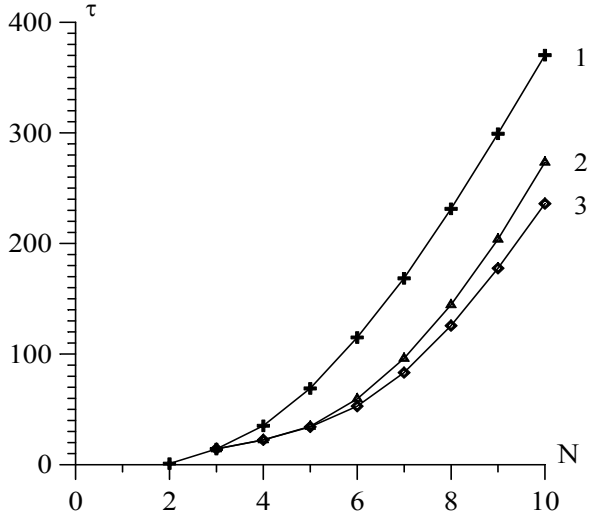
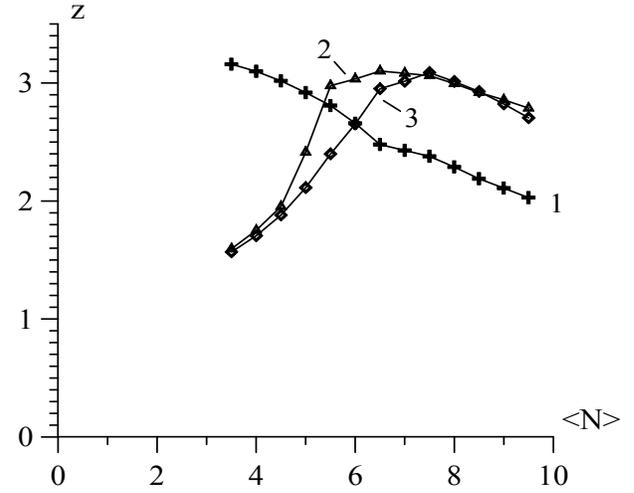

Fig. 3. Dependence of the relaxation time $\tau$ on the number of sites $N$: $T = 0.1$; $H = 0$; 1) $J_2 = 0$; 2) $J_2 = 0.5$; 3) $J_2 = 1$

Fig. 4. Dependence of the critical dynamic exponent $z$ on the average number of sites $<N>$: $T = 0.1$; $H = 0$; 1) $J_2 = 0$; 2) $J_2 = 0.5$; 3) $J_2 = 1$

The dependence of the relaxation time $\tau$ on the number of sites $N$ for a change of the sing of the external magnetic field strength is maintained the same, while in the case where the sign changes for the next nearest neighbor interaction energy, the dependence of $\tau$ on $N$ is quite different (Fig. 5): $J_2$ non-monotonously affects $\tau$ and the latter now tends to depend on parity of the number of sites $N$ (Fig. 5, curve *3*). This is accounted for by the fact that for an even number of sites the total intensity of magnetization of ferromagnetic is equal to zero, as the case should be. While for an odd number of sites the total intensity of magnetization could not be equal to zero, that is to say, the system becomes ferromagnetic. In other words, there is a certain direction due to which non-equilibrium ferrimagnetic reaches the state of equilibrium ferromagnetic faster. With this in mind, we calculated the $z$ exponent separately for the even (initial state – ferromagnetic) and odd (initial state – ferromagnetic) number of sites $N$ (Fig. 6). It is seen in the figure that with increasing the number of sites $<N>$ the value of the $z$ exponent is decreased, as is the case for $J_2 > 0$. Note that there is no monotonous dependence of $z$ on the next nearest neighbor interaction energy $J_2 < 0$. Figures 5 and 6 the interaction energies of the nearest and next nearest neighbors have different signs.

Since for finite systems the phase transition is smeared, we calculated the width of the critical region $\Delta T_c$ [16] as a dimension function of a one-dimensional magnet, which was measured in reduced dimensionless units (Fig. 7). For the number of sites 8, the width of the critical region was $\Delta T_c = 2.3$,

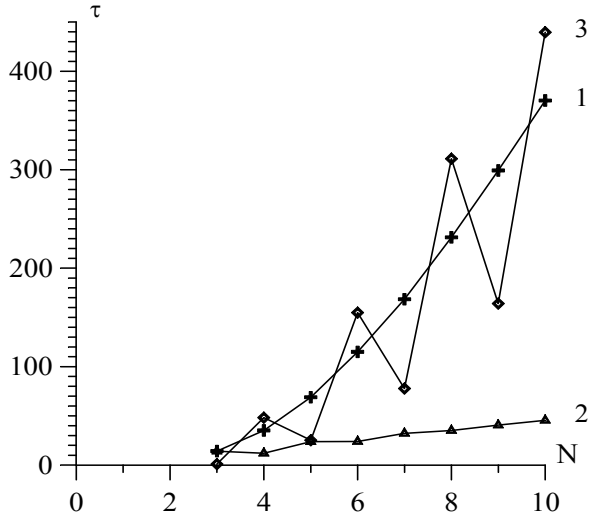
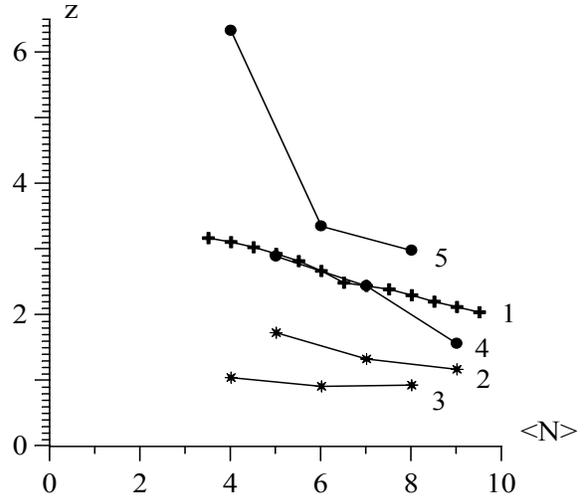

Fig. 5. Dependence of the relaxation time τ on the number of sites $N$: $T = 0.1$; $H = 0$; 1) $J_2 = 0$; 2) $J_2 = -0.5$; 3) $J_2 = -1$

Fig. 6. Dependence of the critical dynamic exponent $z$ on the average number of sites $<N>$: $T = 0.1$; $H = 0$; 1) $J_2 = 0$; 2) $J_2 = -0.5$, $N$ – even; 3) $J_2 = -0.5$, $N$ – odd; 4) $J_2 = -1$, $N$ – even; 5) $J_2 = -1$, $N$ – odd

and for $N = 16$ it was $\Delta T_c = 0.0002$. Then we calculated the critical exponent $z$ at the boundary of the critical region (Fig. 8). Strictly speaking, Eq. (2) is valid for this case only. For the number of sites $<N>$ larger than 10, the exponent values fluctuate around 3, which is larger than for three-dimensional macrosystems (typically $z = 2$).

For finite systems, the phase transition temperature (as points of singularity of thermodynamic functions) is equal to zero. Bearing this in mind, we calculated the critical exponent ν of the correlation length as a function of the number of sites. It increases from 0.016 ($N = 2$) to 0.121 ($N = 16$).

An extrapolation with respect to the inverse dimension resulted in ν = 0.22 in a thermodynamic limit, which is by far smaller than for two- and three-dimensional cases.

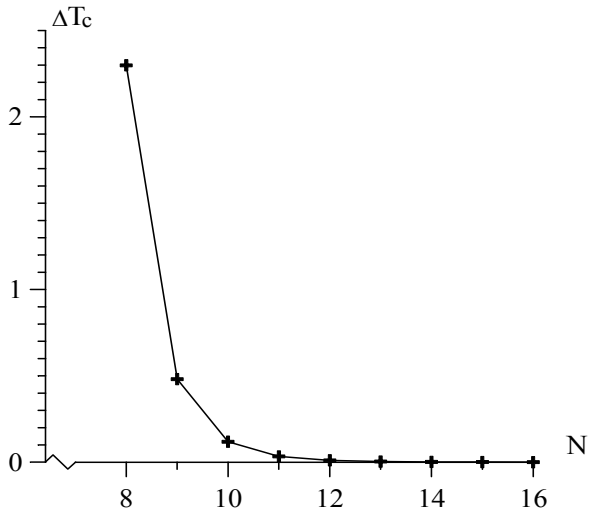
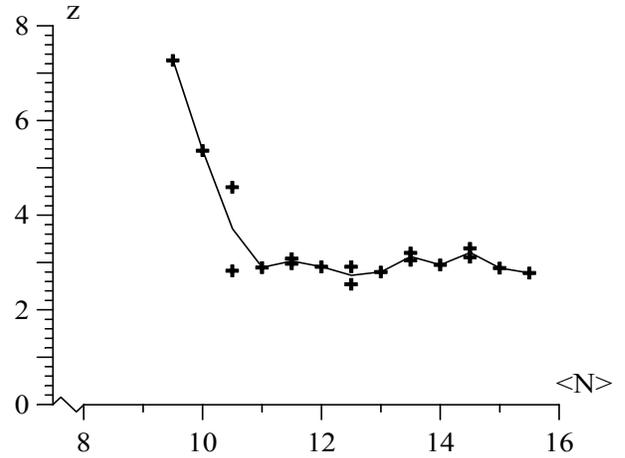

Fig. 7. Dependence of the critical-region width $\Delta T_c$ on the number of sites $N$: $H = 0$; $J_2 = 0$

Fig. 8. Dependence of the dynamic critical exponent on the average number of sites at the critical region boundary: $H = 0$; $J_2 = 0$

In the phase transition region, given that the hypothesis of dynamic scaling is valid, the following relation should fulfill [17]:

$$Y = z\nu, \qquad (4)$$

where $Y$ is the kinetic critical exponent. Its values could be found from the dependence of the relaxation time on temperature $T$ [17] ($T_c = 0$)

$$\tau \propto T^{-Y}. \qquad (5)$$

At the critical region boundary, for $N$ equal to 5, 6, and 7 the exponent $Y > 0$. From the calculations made it follows that the relation $Y = z\nu$ is violated for the average values of parameters. It should be noted, however, that the violations fall within the calculation error. For $N = 6$, $Y = 0.33 \pm 0.15$ (with the relative inaccuracy $\varepsilon = 45\%$), and $z\nu = 0.15 \pm 0.06$ (with the relative inaccuracy $\varepsilon = 40\%$). The calculation error here is a root-mean-square deviation of the values, which is large due to the small size of the model.

Fig. 9. Dependence of the projection of the intensity of magnetization $M_x$ on that of the external magnetic field strength $H_x$: $T = 1$, $J_2 = 0$, $N = 6$ (*a*), $N = 10$ (*b*)

We calculated the hysteresis of the projection of the intensity of magnetization (Fig. 9) and investigated the evolution of the hysteresis loop with respect to the number of cycles for the number of sites $N = 6$ and 10. As the number of sites increases, the loop no long depends on the number of cycles, i.e., it becomes steady state for a larger number of cycles; thus, the degree of the process nonequilibrium is higher. The shape of the calculated hysteresis loop qualitatively agrees with the macroscopic experimental data [2, 18].

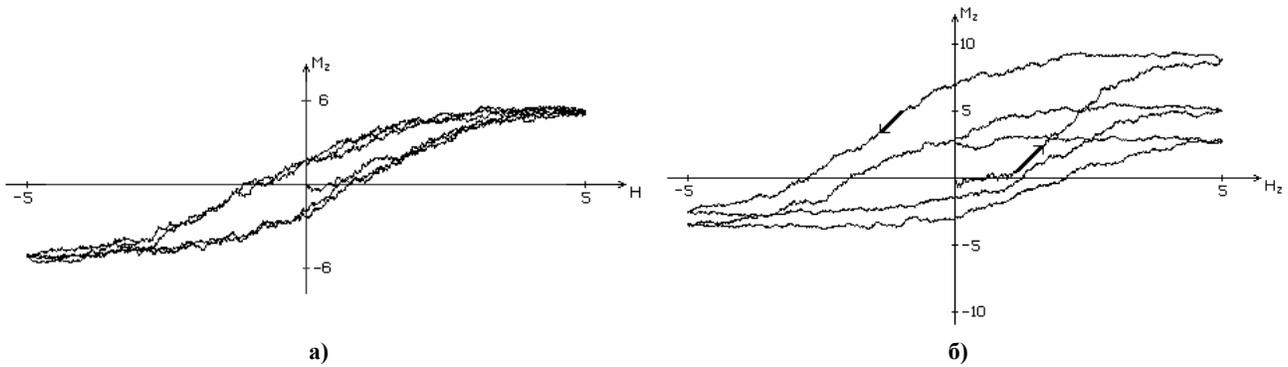

а)   б)

Fig. 9. Dependence of the projection of the intensity of magnetization $M_x$ on that of the external magnetic field strength $H_x$: $T = 1$, $J_2 = 0$, a) $N = 6$; б) $N = 10$

To conclude, the following should be summarized:
1. For small one-dimensional magnet undergoing a non-equilibrium antiferromagnetic (ferromagnetic) – ferromagnetic transition, estimates of the dynamic critical exponent $z$ have been obtained, including those for an external magnetic field and taking into account the interactions of the next nearest spins at a constant reduced temperature of 0.1. The maximum value of $z$ was found to be 3.5.

2. It has been shown for a comparatively large system ($17 > N > 11$) that at the critical-region boundary the dynamic critical exponent values fluctuate about 3, which is lager than for three-dimensional macrosystems (typically $z = 2$). According to our results, the maximum value of the dynamic critical exponent $z$ is equal to 22 ($N < 6$).

3. For small magnet, a hysteresis of the dependence of the projection of magnetization intensity $M_x$ on the external field strength $H_x$, has been calculated, with the field cyclically changing. The evolution of the hysteresis loop has been investigated with respect to time. It should be noted that for the magnet in question, the shape of the calculated hysteresis loop qualitatively agrees with the macroscopic experimental data for three-dimensional systems.

4. It has been demonstrated that for small magnet the hypothesis of dynamic scaling is considerably violated for the average parameter values.

Thus, there is ample evidence to expect that under non-equilibrium phase transitions small one-dimensional nanosized magnets would behave in a manner qualitatively different than that of macroscopic magnetic systems.